\documentclass[
    prd, twocolumn, superscriptaddress, nofootinbib, amsmath, amssymb,
    aps, floatfix, preprintnumbers, 10pt
]{revtex4-2}
\usepackage[utf8]{inputenc}
\usepackage{setspace}
\usepackage[dvips]{graphicx}
\usepackage[dvipsnames]{xcolor}
\definecolor{CiteBlue}{RGB}{45,52,151}
\usepackage[
    colorlinks=true,
    linkcolor=CiteBlue,
    urlcolor=CiteBlue,
    citecolor=CiteBlue
]{hyperref}
\usepackage{aas_macros}
\usepackage[capitalise]{cleveref}
\crefname{section}{Sec.}{Secs.}
\usepackage{times}

\newcommand{\refcite}[1]{Ref.~\cite{#1}}
\newcommand{\refscite}[1]{Refs.~\cite{#1}}

\usepackage[version=4]{mhchem}
\usepackage{siunitx}
\DeclareSIUnit{\year}{yr}
\DeclareSIUnit{\au}{AU}
\DeclareSIUnit{\parsec}{pc}

\usepackage{bm}

\newcommand{\du}{\mathrm{d}}
\newcommand{\pbh}{\mathrm{PBH}}
\newcommand{\fpbh}{f_\pbh}
\newcommand{\fdm}{\mathcal{F}_\dm}
\newcommand{\cl}{\mathrm{cl}}
\newcommand{\bin}{\mathrm{bin}}
\newcommand{\dm}{\mathrm{DM}}
\newcommand{\vorb}{v_{\mathrm{orb}}}
\newcommand{\psma}{a_\phi}
\newcommand{\parallax}{\Pi}
\newcommand{\Gaia}{\textit{Gaia}}

\begin{document}
\title{Dancing with invisible partners: Three-body exchanges with primordial black holes}
\preprint{MIT-CTP/5743}

\author{Badal Bhalla}
\email{badalbhalla@ou.edu}
\affiliation{Homer L. Dodge Department of Physics and Astronomy, University of Oklahoma, Norman, OK 73019, USA}

\author{Benjamin V. Lehmann}
\email{benvlehmann@gmail.com}
\affiliation{Center for Theoretical Physics, Massachusetts Institute of Technology, Cambridge, MA 02139, USA}

\author{Kuver Sinha}
\email{kuver.sinha@ou.edu}
\affiliation{Homer L. Dodge Department of Physics and Astronomy, University of Oklahoma, Norman, OK 73019, USA}

\author{Tao Xu}
\email{tao.xu@ou.edu}
\affiliation{Homer L. Dodge Department of Physics and Astronomy, University of Oklahoma, Norman, OK 73019, USA}

\begin{abstract}
The abundance of massive primordial black holes has historically been constrained by dynamical probes. Since these objects can participate in hard few-body scattering processes, they can readily transfer energy to stellar systems, and, in particular, can disrupt wide binaries. However, disruption is not the only possible outcome of such few-body processes. Primordial black holes could also participate in exchange processes, in which one component of a binary system is ejected and replaced by the black hole itself. In this case, the remaining object in the binary would dynamically appear to have an invisible companion. We study the rate of exchange processes for primordial black holes as a component of dark matter and evaluate possible mechanisms for detecting such binaries. We find that many such binaries plausibly exist in the Solar neighborhood, and show that this process can account for observed binary systems whose properties run counter to the predictions of isolated binary evolution.
\end{abstract}

\maketitle

\section{Introduction}
\label{sec:introduction}
The microphysical identity of cosmological dark matter (DM) remains unknown. Given the difficulty in accounting for cosmological observables by modifications to gravity, the DM is thought to be made up of a new particle species beyond the Standard Model. Another possibility, however, is that part or all of the DM is in the form of primordial black holes (PBHs) produced at early times~\cite{Zeldovich:1967lct,Hawking:1971ei,Carr:1974nx,Carr:2016drx,Carr:2020xqk,Green:2020jor,Carr:2024nlv,Green:2024bam}. Given the null results of many particle DM searches, and an abundance of new observational probes of black holes (BHs), interest in PBHs as a DM candidate has surged over the last decade.

PBHs as DM are constrained by a multitude of observables over many decades in mass~\cite{Carr:2016drx,Carr:2020xqk,Green:2020jor}. It is generally agreed that PBHs cannot account for all of DM outside a narrow window of masses, around \num{e17}--\qty{e23}{\gram}~\cite{Sugiyama:2019dgt,Montero-Camacho:2019jte,Smyth:2019whb,Gorton:2024cdm}. Still, PBHs may account for up to 10\% of the DM abundance in other key mass ranges, such as the planetary mass scale, \num{e-7}--\qty{e-3}{M_\odot}. As such, there has been a robust community effort to develop new observables that might further constrain or produce direct evidence of such objects.

One of the key approaches relies on the dynamics of these objects~\cite{Brandt:2016aco,Koushiappas:2017chw,Zhu:2017plg,Stegmann:2019wyz,Graham:2023unf,1985ApJ...290...15B,Yoo:2003fr,Quinn:2009zg,Allen:2014tla,Monroy-Rodriguez:2014ula,Tyler:2022rxi}. On large scales, PBHs are indistinguishable from a fluidlike background of particle DM\@. But on sufficiently small scales, the inhomogeneity of the PBH distribution relative to particle DM leads to unique phenomenology. In particular, such objects can lose their kinetic energy via interactions with stellar systems. One simple way to understand this is via thermodynamics~\cite{Graham:2023unf}: since PBH velocities are determined by the global properties of a galactic halo, the effective temperature that characterizes the PBHs' phase space distribution scales with the PBH mass. Since stars have comparable velocities, this implies that when the PBH mass is significantly larger than the stellar masses, the PBH fluid is hotter, and tends to heat the stellar fluid.

Beyond these considerations which apply to the stellar distribution as a whole, PBHs can also participate in few-body encounters with binary systems, with behavior that is radically different from particle DM\@. In particular, a single three-body encounter between a PBH and a binary system can transfer enough energy to substantially widen or even unbind the binary. This means that the distribution of binary widths can be used to constrain the population of would-be perturbers such as PBHs, and this exact process has historically provided some of the most important constraints on massive PBHs with $M_\pbh \gtrsim \qty{100}{M_\odot}$~\cite{1985ApJ...290...15B,Yoo:2003fr,Quinn:2009zg,Allen:2014tla,Monroy-Rodriguez:2014ula,Tyler:2022rxi}.

But the exchange of energy is not the only possibility in three-body encounters, which have been extensively studied in other contexts~\cite{Heggie:1975rcz,1975AJ.....80..809H,Quinlan:1996vp,2006tbp..book.....V}. Consider the scattering of three bodies 1, 2, and 3, where 1 and 2 are initially bound and 3 is initially free. In general, there are five possible outcomes:
\begin{enumerate}
    \item Hardening: objects 1 and 2 remain bound, and they give up energy to object 3, causing their separation to decrease. Object 3 remains free.
    \item Softening: object 3 transfers energy to the 1-2 system, causing their separation to increase, but they remain bound. Object 3 remains free.
    \item Disruption: object 3 transfers so much energy to the 1-2 system that the two objects are unbound. All objects become free.
    \item Capture: objects 1 and 2 remain bound, and now object 3 becomes bound as well, forming a triple system. In the process, 3 donates energy to the 1-2 system, softening the binary.
    \item Exchange: object 3 transfers enough energy to object 1 to unbind it from 2, but in the process, 3 loses enough energy that it becomes bound to 2. The result is a bound 2-3 system with object 1 free.
\end{enumerate}
Softening and disruption by PBHs have been considered extensively in the literature~\cite{1985ApJ...290...15B,Yoo:2003fr,Quinn:2009zg,Allen:2014tla,Monroy-Rodriguez:2014ula,Tyler:2022rxi}, and PBH capture has also been treated by \refcite{Lehmann:2022vdt}. For reasons that we will explain in the next section, hardening processes involving PBHs are strongly suppressed in all realistic systems.

This leaves one untapped possibility: exchange processes, where a PBH ejects one component of a binary system and forms a new binary with the other component. The end result of this process is a single visible object which has dynamics characteristic of a binary component. That is, such an object appears to be dancing with an invisible partner, with potentially observable implications. Here we investigate the possibility that such a population of invisible partners exists in our Galaxy, and we survey the observational approaches that might be used to establish their presence.

Identifying a binary composed of a PBH and a visible object would require a means of ruling out an astrophysical origin for the BH component of the binary. For subsolar-mass BH components, this only requires establishing that the object is in fact a BH\@. However, we will see that the most promising prospects for detecting PBH exchange are in stellar binaries, at PBH masses comparable to astrophysical BH masses. Here, a primordial origin can be established by rejecting standard astrophysical formation channels. Interestingly, the \Gaia{} mission~\cite{Gaia:2016zol} has detected several BH-star binary systems that pose a challenge to standard astrophysical binary formation models, with orbital periods too long and mass ratios too large to have been produced from isolated binaries that underwent a common-envelope phase. As we will demonstrate, PBH exchange is potentially viable as an origin for these systems, meaning that BH components in BH-star binaries may in fact be identifiable as PBHs.

The remainder of this article is organized as follows. In \cref{sec:exchange-processes}, we review the generalities of the three-body exchange process, and detail our calculation of exchange rates. In \cref{sec:candidate-binary-systems}, we systematically consider the classes of binary systems that might be amenable to the exchange process. In \cref{sec:gaia}, we examine the possibility that observed BH-star binaries may form via PBH exchange. We discuss our findings and conclude in \cref{sec:discussion}.

\section{Three-body processes}
\label{sec:exchange-processes}

\begin{figure}\centering
    \includegraphics[width=\columnwidth]{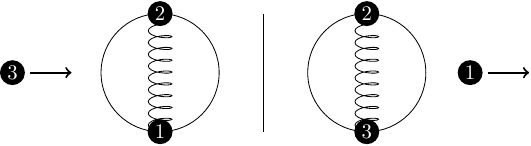}
    \caption{Summary of notation. In the initial state, objects 1 and 2 are bound, and object 3 is free. In the final state, objects 2 and 3 are bound, and object 1 is free.
    }\label{fig:exchange-diagram}
\end{figure}

Three-body dynamics is a notoriously complicated subject, with a paucity of fully general analytical results. Still, for the exchange processes that will be of interest to us, we can leverage approximations that have been developed in the astrophysics literature and validated against suites of numerical experiments. We now summarize these results as they apply to PBH exchanges with binaries. In this section, and in the rest of this work, we use the following notation: we take the three objects to have masses $m_1$, $m_2$, and $m_3$. We use the label 3 to denote the free object in the initial state, that is, the PBH\@. We use the label 2 to denote the binary component that forms a new binary with the PBH in the final state, and we use the label 1 to denote the object that is ejected in the encounter. This is summarized in \cref{fig:exchange-diagram}. We further define
\begin{equation}
    m_{ij} \equiv m_i + m_j,
    \qquad
    m_{123} \equiv m_1 + m_2 + m_3,
\end{equation}
and we write $\mu_1 \equiv m_1/m_{12}$ and $\mu_2 \equiv m_3/m_{123}$. We denote the semimajor axis of the initial binary by $a$, and denote the velocity of the incoming object at infinity by $v_3$, or equivalently by $v_\pbh$ when our discussion is specific to PBHs.

One of the major complications in the study of three-body processes is the sensitivity to initial conditions: slight variations in the parameters of the system can lead to wildly different outcomes. However, we are mainly interested in the rate of certain three-body processes on average, over a large number of systems. This is considerably simpler to analyze, since it is often possible to analytically describe the statistics of outcomes of three-body encounters. In particular, the seminal work of Heggie~\cite{Heggie:1975rcz} and Hills~\cite{1975AJ.....80..809H} provides a simple framework for estimating the rates of various processes.

As regards energy transfer, their results can be summarized in Heggie's law, which states that in a three-body encounter, on average, hard binaries become harder (or ``shrink'') and soft binaries become softer (or ``expand''). For these purposes, a binary system is considered hard if its binding energy $U_{12}$ is much greater than the average kinetic energy $\frac12m_3\langle v_3^2\rangle$ of the third object. Conversely, a binary system is classified as soft if $U_{12} \ll \frac12m_3\langle v_3^2\rangle$~\cite{2008gady.book.....B}. The disruption of stellar binaries as a strategy to constrain exotic objects has a long history: it has been effective in setting bounds on MACHOs~\cite{1985ApJ...290...15B,Yoo:2003fr,Quinn:2009zg,Allen:2014tla,Monroy-Rodriguez:2014ula,Tyler:2022rxi} and DM substructure~\cite{Penarrubia:2010pa, Ramirez:2022mys, Penarrubia:2016ltr}. Our goal in this work is to explore the other possible outcomes of PBH encounters.

One possibility is to consider encounters with very light PBHs, which are poorly constrained. Instead of disrupting (or ``softening'') the binary with a heavy perturber, $m_3 \gg m_1,m_2$, one can consider tightening (or ``hardening'') the binary with a light perturber, $m_3 \ll m_1, m_2$. However, the transition between the hard and soft binary regimes is not well defined, and this leads to significant complications for light DM-like perturbers. According to numerical experiments by Hills~\cite{1990AJ.....99..979H}, perturbers with a velocity larger than the orbital speed of the binary will tend to soften the binary, while slower perturbers will tend to harden it, irrespective of their masses. This disagrees with the earlier definition for the case of a fast, low-mass perturber, which is exactly the regime of interest for hardening by low-mass PBHs. Quinlan~\cite{Quinlan:1996vp} suggests that a binary in a sea of low-mass perturbers must not be considered hard unless its orbital speed exceeds the perturber velocity dispersion by a factor that proportional to $\left( 1 + m_1/m_2 \right)\null^{1/2}$, where $m_1\geq m_2$. This mass dependence corresponds to that of a critical velocity $w(m_1, m_2)$ for capture. Perturbers with velocity below $w$ can be captured by the binary, altering the statistics of the outcomes of close encounters.

The cutoff imposed by the critical velocity is fatal to the prospect of observing hardening by light PBHs, as we explain in detail in \cref{sec:hardening}. Briefly, the rate of hardening scales as $\du a/\du t \propto a^2$, so the hardening effect is only discernible at large binary separations, $a \gtrsim \qty{e5}{\au}$. Thus, one is immediately locked into binaries with very wide separations, where data is sparse and observational prospects are already challenging. But since the orbital velocity of the binary decreases with increasing separation, only a vanishingly small fraction of PBHs would have a velocity lower than the binary's orbital speed, unless we assume very cold DM systems.

Thus the only interesting possibility that remains is exchange. We now consider the conditions needed for exchange to take place. An exchange requires that the three objects do not all escape to infinity after the encounter, which requires that the initial velocity of the PBH is small. This translates to another maximum velocity $v_{\max}$, different from $w$, below which exchange is possible. In the rest of this work, we follow \refcite{Heggie:1996bs} and set $v_{\max}$ to be the velocity at which the total energy of the three-body system is zero in the rest frame of their barycenter, i.e.,
\begin{equation}
    v_{\max} = \sqrt{\frac{Gm_1m_2m_{123}}{m_{12}m_3a}}
    .
\end{equation}
Thus, when computing the exchange cross section, we require that $v_\pbh < v_{\max}$. For the purposes of exchange, this is the condition for the binary to be hard with respect to the perturber. Quantitatively, this corresponds to the condition that
\begin{equation}
    \label{eq:hardness-condition}
    v_3 < v_{\max} = (\qty{30}{\kilo\meter/\second})
    \left( \frac{a}{\qty{1}{\au}} \right)^{-1/2}
    \left( \frac{\frac{m_1m_2}{m_3}\frac{m_{123}}{m_{12}}}{{\qty{}{M_\odot}}} \right)^{1/2}
    .
\end{equation}
For the case of a hard binary, we compute the exchange cross section $\Sigma$ following \refcite{Heggie:1996bs}, which gives
\begin{multline}
    \label{eq:exchange-cross-section}
    \Sigma(v_3) =
        \qty{13.9}{\au^2} \times
        \left(\frac{a}{\qty{1}{\au}}\right)
        \left(\frac{v_3}{\qty{10}{\kilo\meter/\second}}\right)^{-2}
        \left(\frac{m_3}{m_{13}}\right)^{5/2}
        \\\times
        \frac{m_3}{m_{12}}
        \frac{
            \left(m_{12}^4m_{23}m_{123}\right)^{1/6}
        }{\qty{1}{M_\odot}}
        \exp\Bigl\{
            3.70 + 7.49\mu_1
            - 1.89\mu_2
            \\
            - 15.49\mu_1^2 - 2.93\mu_1\mu_2 
            - 2.92\mu_2^2
            + 3.07\mu_1^3
            \\
            + 13.15\mu_1^2\mu_2
            - 5.23\mu_1\mu_2^2
            + 3.12\mu_2^3
        \Bigr\}
        .
\end{multline}
This expression is a fit to numerical experiments conducted with $v_\pbh = 0.1v_{\max}$. We assume the validity of this expression for all velocities $v<v_{\max}$, and also define the velocity-independent quantity $\Sigma_0\equiv \Sigma\times v_3^2$. We take all binaries to be circular throughout this work.

Typically, the rate of exchanges will be dominated by the low-velocity tail of the PBH distribution. We take the PBH speed distribution to be that of Galactic DM, such that the Probability Density Function (PDF) is given by

\begin{equation}
    \label{eq:vdf}
    \fdm(v_\pbh) = \sqrt{\frac2\pi}\frac{v_\pbh^2}{\sigma_\dm^3}
        \exp\left[-\frac{v_\pbh^2}{2\sigma_\dm^2} \right]
    .
\end{equation}
Then the overall exchange rate per binary is then given by
\begin{multline}
    \label{eq:rate-per-binary}
    \Gamma =
    n_\pbh^{v<v_{\max}}\langle \Sigma(v_3) v_3\rangle
    \\
    = n_\pbh \int_0^{v_{\max}}\du v_3\,
        \fdm(v_3)\Sigma(v_3) v_3
    ,
\end{multline}
where $n_\pbh$ is the total number density of PBHs, $n_\pbh^{v<v_{\max}}$ is the number density of PBHs with $v_\pbh < v_{\max}$, and $\langle\cdot\rangle$ denotes the average with respect to the distribution of PBH velocities $\fdm(v_3)$ subject to the restriction $v_3<v_{\max}$. For $v_{\max} \ll \sigma_\dm$, we expand $\fdm$ to find
\begin{equation}
    \label{eq:approximate-rate}
    \Gamma
    \simeq n_\pbh\Sigma_0
        \int_0^{v_{\max}}\du v_3\,
        \sqrt{\frac2\pi}\frac{v_3}{\sigma_\dm^3}
    = \frac{n_\pbh\Sigma_0v_{\max}^2}{\sqrt{2\pi}\sigma_\dm^3}
    .
\end{equation}
In particular, since $\Sigma_0 \propto a$ and $v_{\max}\propto a^{-1/2}$, this implies that $\Gamma$ is approximately independent of $a$. For fixed masses, this approximation breaks down at small separations, where $v_{\max}/\sigma_\dm$ becomes large. However, we will see that \cref{eq:approximate-rate} remains valid throughout most of the parameter space of interest in this work.

\begin{figure}\centering
    \includegraphics[width=\columnwidth]{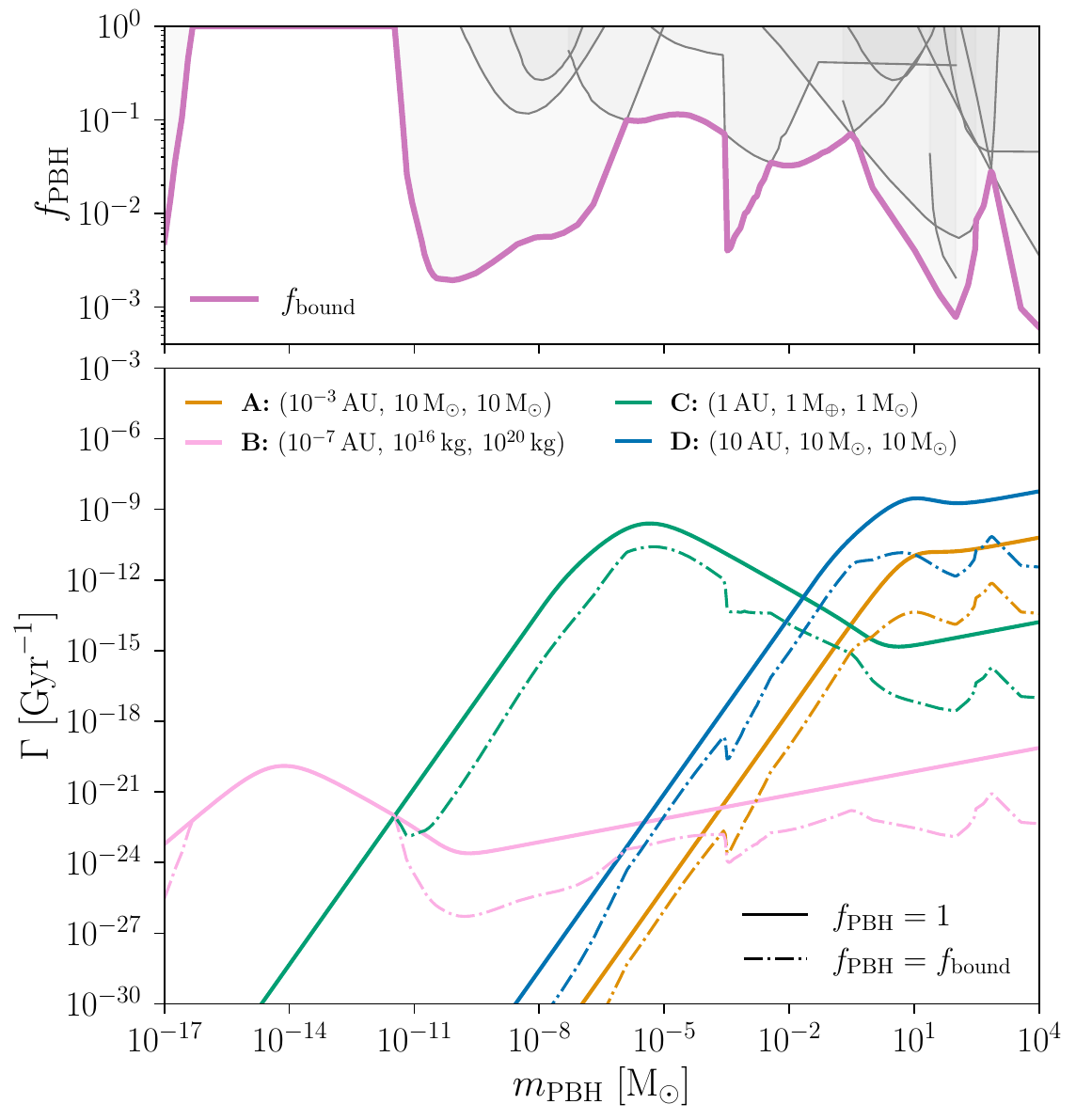}
    \caption{Bottom: exchange rate as a function of PBH mass for some typical binary systems. The DM dispersion is taken to be $\sigma_\dm = \qty{220}{\kilo\meter/\second}$, and binary parameters are given in the legend in the sequence $(a, m_1, m_2)$. Curves A, B, C, and D are typical of close compact object binaries, asteroid binaries, exoplanetary systems, and stellar binaries, respectively. Solid curves assume that $\fpbh = 1$, and dot-dashed lines take $\fpbh$ to saturate observational constraints, i.e., $\fpbh = f_{\mathrm{bound}}$. Top: observational constraints on $\fpbh$~\cite{Ali-Haimoud:2016mbv,EROS-2:2006ryy,Carr:2009jm,Lehmann:2018ejc,Jung:2017flg,Brandt:2016aco,Griest:2013aaa,Macho:2000nvd,Niikura:2017zjd,Dror:2019twh,Chen:2019irf,LIGOScientific:2019kan,Kavanagh:2018ggo}\footnote{Constraints are taken from the \href{https://github.com/bradkav/PBHbounds}{PBHBounds repository}~\cite{2019zndo...3538999K}.}. The purple curve shows the maximum allowed value of $\fpbh$ at each mass, $f_{\mathrm{bound}}$, assuming a monochromatic mass function.}
    \label{fig:exchange-rate-comparison}
\end{figure}

\begin{figure*}\centering
    \includegraphics[width=\textwidth]{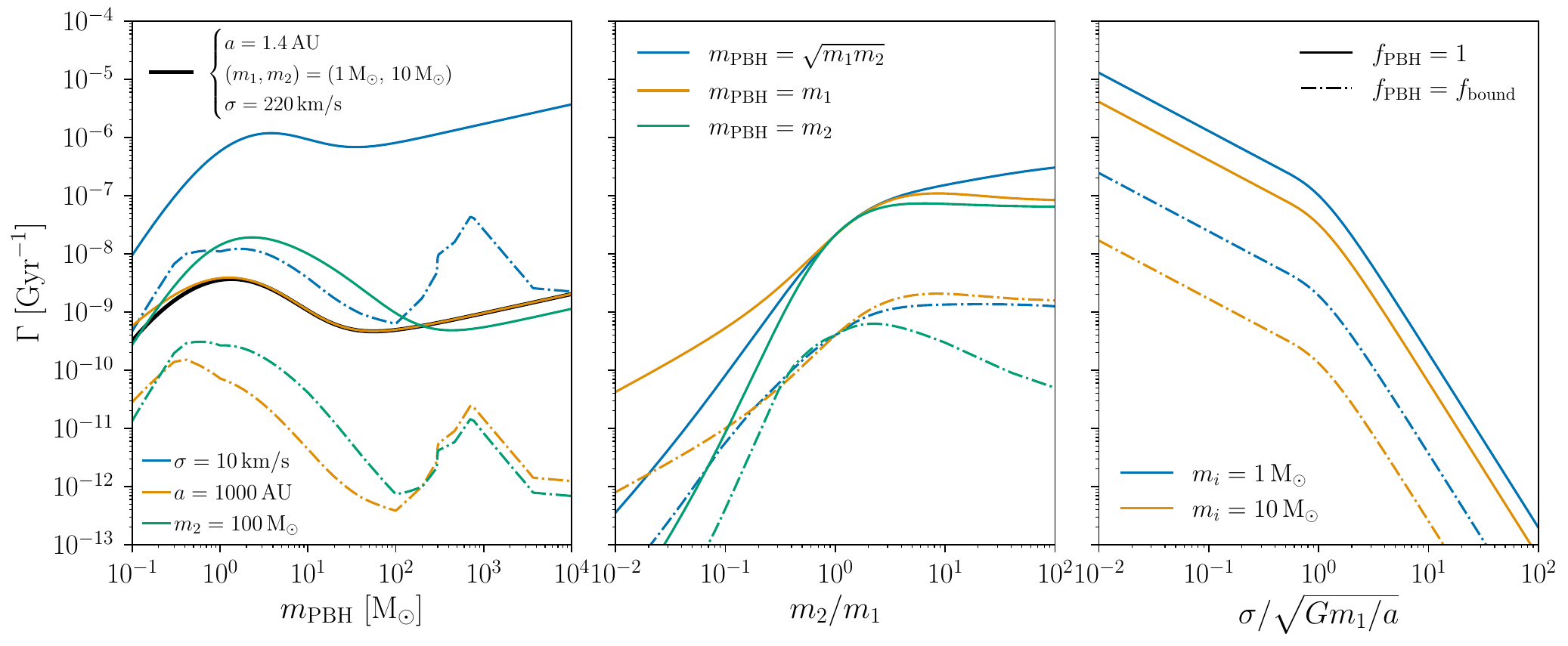}
    \caption{Exchange rate as a function of PBH mass for typical stellar binary systems. Solid curves assume that all of the DM is composed of PBHs, with a DM density of \qty{0.4}{\giga\electronvolt/\centi\meter^3}. Dot-dashed curves assume that $\fpbh$ saturates observational bounds. (See \cref{fig:exchange-rate-comparison}.) Left: exchange rate as a function of PBH mass. Center: exchange rate as a function of the mass ratio of the bound object to the ejected object. Note that the bound on the PBH abundance is dependent on the choice of $m_\pbh$, meaning that the ordering of the dot-dashed curves can vary with $m_2/m_1$. Right: exchange rate as a function of the DM velocity dispersion in ratio to orbital velocity scale of the binary, with all three objects taken to have equal mass.}
    \label{fig:stellar-binary-exchange-rate}
\end{figure*}

We show the exchange rate for a variety of different binary systems in \cref{fig:exchange-rate-comparison}. We assume that PBHs account for a fraction $\fpbh\leq1$ of the DM density with a monochromatic mass distribution, and we show results corresponding both to $\fpbh = 1$ (solid curves) and to $\fpbh$ saturating observational constraints at each mass (dot-dashed curves). Note that each of these observational bounds has been contested, as summarized by \refscite{Carr:2020gox,Carr:2020xqk,Green:2020jor,Carr:2023tpt}. In particular, we omit the nominal constraint from the OGLE experiment, which has been interpreted to show a preference for a population of Earth-mass PBHs over the null hypothesis of $\fpbh = 0$~\cite{Niikura:2017zjd}. (However, see \refscite{Mroz:2024mse,Esteban-Gutierrez:2023qcz,Gorton:2022fyb}.)

\Cref{fig:stellar-binary-exchange-rate} shows the exchange rate and its functional dependences for benchmark parameters typical of stellar binaries. As is clear from the figure, the rate generally peaks when $m_1 \simeq m_3$, i.e., when the mass of the PBH matches the mass of the object to be ejected. These curves in the left and center panels take the PBH velocity dispersion to be $\sigma_\dm = \qty{220}{\kilo\meter/\second}$, but per \cref{eq:approximate-rate}, the rate is much higher in cold systems with smaller dispersions, as shown in the right panel. While one might expect the rate to decrease with $m_2$, the mass of the secondary, the opposite is true, as demonstrated in the center panel. Instead, increasing the mass of the secondary increases the gravitational focusing effect, and enhances the effective cross-section for PBHs to scatter with the binary system.

Counterintuitively, the exchange rate also increases with the PBH mass, even while holding the PBH density constant. When $m_3 \gg m_1, m_2$, the critical velocity $v_{\max}$ becomes independent of $m_3$. Then all $m_3$ dependence in the cross section is confined to $\Sigma_0$, which goes as $\Sigma_0 \propto (m_3^4/m_{12})^{1/3}\qty{}{M_\odot^{-1}}$. Since $n_\pbh \propto m_3^{-1}$ at fixed abundance, the overall rate scales with $\Gamma \propto m_3^{1/3}$. When the PBH is much more massive than the binary, the kinetic energy of the PBH is large compared to the potential energy of the binary in the rest frame of the binary. This naively facilitates disruption. However, the typical velocity imparted to the binary components is set by $v_\pbh \ll v_{\max}$, and in the limit of a massive perturber this implies that at most one object can escape to infinity. This is shown in detail in Sec.~3.1 of \refcite{Heggie:1996bs}. Thus, exchange is inevitable: after an encounter with significant energy transfer, one of the components of the original binary system is typically captured by the PBH, even if the other escapes. Thus, captures and exchanges dominate over disruption in this regime.

There is nothing nonphysical about the rate growing without bound as $m_3 \to \infty$: in this limit, the number of PBHs eventually shrinks to just one with the mass of the entire Galactic halo. Then all binaries of interest must immediately be bound to the PBH simply by virtue of being bound to the halo itself. However, at much lower masses, a more pragmatic cutoff is imposed by the radius of PBH, which exceeds \qty{1}{\au} for $m_3 \gtrsim \qty{5e7}{M_\odot}$. When the radius of the PBH exceeds the separation of the binary system in question, it is no longer appropriate to treat the PBH as a point particle in Newtonian gravity. At any rate, since other dynamical constraints become severe at such large masses, we restrict our attention to $m_3 < \qty{e4}{M_\odot}$ in this work.

For all the binaries shown, the timescale for the exchange process is above the Hubble time. However, given the number of binary systems in the Milky Way, it is still plausible that an exchange process has occurred in a large number of them. In the next section, we consider the specific types of binary systems that are best suited to undergo exchange with a PBH\@.

\section{Candidate binary systems}
\label{sec:candidate-binary-systems}
The exchange process described in \cref{sec:exchange-processes} can take place across an enormously wide range of binary systems, with widely varying parameters. In particular, the separations span an enormous range: the widest observed stellar binaries have separations of order \qty{e5}{\au}, while asteroid binaries can be as close as \qty{e-8}{\au}, or just a few kilometers. The masses of binary components---and, of course, of hypothetical PBHs---span many orders of magnitude as well. In this section, we outline the classes of observable binary systems, and we briefly evaluate the extent to which each type of system might be amenable to exchanges with PBHs.

\subsection{Compact object binaries}
Compact object binaries are a very unusual class of objects for our present purposes: they inhabit unique regions of binary parameter space, and also offer unique observable signatures. Binaries with at least one black hole component are most readily observed via gravitational waves from their mergers. In order to merge within a Hubble time, the binary must already be quite close, meaning that the maximum velocity $v_{\max}$ for the exchange process, given in \cref{eq:hardness-condition}, is significantly increased. In particular, observed BH-BH binaries are thought to form from a common envelope phase, when both objects are enclosed within the radius of a single giant star, typically well below \qty{1}{\au}. This implies that there is very little suppression of the exchange rate from requiring hardness of the binary with respect to the perturber, and also renders our estimate in \cref{eq:approximate-rate} inadequate.

However, there are a number of inherent difficulties in studying a PBH population via exchanges in compact object binaries. Firstly, in such a system, it would likely be impossible to establish that the exchanged black hole had a primordial origin rather than a astrophysical origin. Since the exchange is maximal when the PBH mass is at least as large as the masses of the binary components, the best prospects for PBH exchange are at black hole masses that are already produced by stellar evolution. The BH spin parameter measured in gravitational waves from a merger could suggest a primordial origin, since PBHs are known to have negligible spins when formed in a radiation-dominated universe~\cite{Chiba:2017rvs, Mirbabayi:2019uph,DeLuca:2019buf}, but making such a statement robustly would require a large number of events.

Secondly, since the main observable signatures of compact object binaries arise from their mergers, it is necessary to consider the impact of the exchange process on the time to merger. Since binaries that will eventually merge must already be very hard, other three-body processes have little impact on their evolution. But exchanges transfer a large amount of energy from the perturber to the ejected object, so the time to merger between the components of the new binary system can be significantly different from the time to merger of the original binary. As such, even given the exchange rate in these binaries, it is not trivial to estimate the observed merger rate of systems that have undergone an exchange with a PBH\@.

Still, the biggest challenge in using these systems to probe PBHs is that merger signatures are transient events. This means that in any given time period, the number of observable systems can be very small even if the total number of systems is large. In principle, gravitational wave observatories are sensitive to mergers at cosmological distances, encompassing a vast number of sources, but very few of these binaries will merge during the observing period.

\Cref{fig:exchange-rate-comparison} indicates that the exchange rate per binary is at most $\mathcal O(\qty{e-20}{\per\year})$. Under the most optimistic circumstances, let us consider all mergers with redshifts $z<1$ to be detectable. Certainly some merger events are detectable at such distances---for example, the merger that produced the gravitational wave event GW190521 took place at an estimated redshift of $z\approx 0.8$~\cite{LIGOScientific:2020ufj}. The Milky Way is estimated to contain $\mathcal O(\num{e6})$ BH binaries~\cite{Lamberts:2018cge}, and there are approximately \num{e9} galaxies with $z < 1$~\cite{2016ApJ...830...83C}. Assuming that all galaxies have as many BH binaries as the Milky Way leads to a significant overestimate of the population. If binaries merge immediately after exchanges, then the overall rate of mergers is still at most \qty{e-5}{\per\year}. Thus, it is extremely unlikely that the current generation of gravitational wave detectors would be capable of detecting any binaries have undergone an exchange with a PBH\@. Even future facilities such as Cosmic Explorer or Einstein Telescope will not change this conclusion: the total number of galaxies in the observable Universe is no larger than $\mathcal O(\num{e12})$~\cite{2016ApJ...830...83C,Lauer:2020qwk}, so the total rate of exchanges is well below $\qty{e-2}{\per\year}$.

\begin{figure*}\centering
    \includegraphics[width=\textwidth]{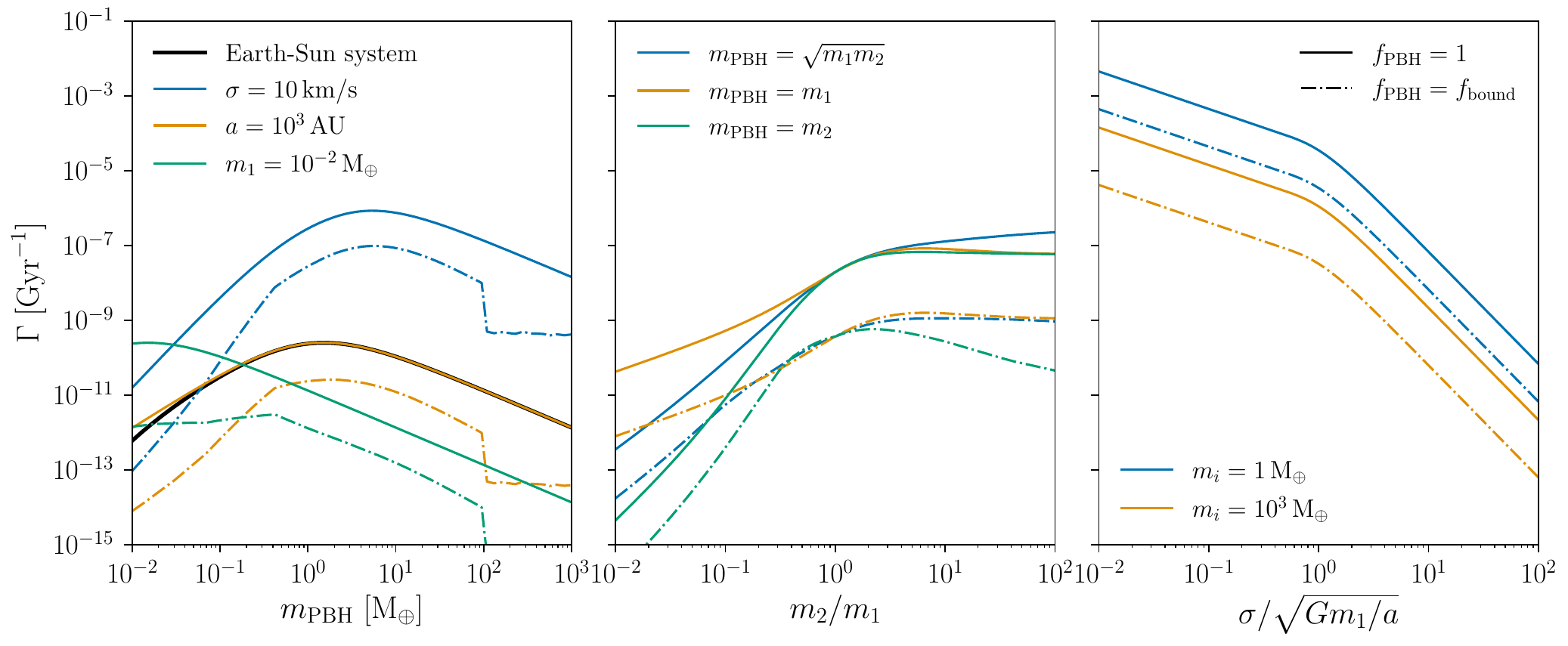}
    \caption{Exchange rate as a function of PBH mass for typical exoplanetary systems. Solid curves assume that all of the DM is composed of PBHs, with a DM density of \qty{0.4}{\giga\electronvolt/\centi\meter^3}. Dot-dashed curves assume that $\fpbh$ saturates observational bounds. (See \cref{fig:exchange-rate-comparison}.) Left: exchange rate as a function of PBH mass. Center: exchange rate as a function of the mass ratio of the bound object to the ejected object. Right: exchange rate as a function of the DM dispersion velocity in ratio to orbital velocity scale of the binary, with all three objects taken to have equal mass.}
    \label{fig:exchange-rate-planet}
\end{figure*}

\subsection{Asteroid binaries}
Asteroid binaries provide an interesting laboratory for dynamics on a unique set of scales, and observational measurements of their dynamics are accumulating rapidly. The list of observed asteroid binaries has expanded significantly in recent years, and now numbers 542: 98 near-Earth asteroids, 34 Mars-crossing asteroids,
262 main-belt asteroids, 8 Jupiter Trojan asteroids, and 140 trans-Neptunian objects~\cite{2015aste.book..355M, 2016Icar..267..267P, 2019Icar..334...62G, 2023SSRv..219...59N, 2015aste.book..375W}.
The binary separations range from $\mathcal{O}(1)$ km to ultra-wide trans-Neptunian binaries, which can have separation  $\mathcal{O}(\num{e5})\,\qty{}{\kilo\meter}$. The relative radius of the secondary component of the binary system compared to the primary component can range from 4--50\% for near-Earth binaries to ${\sim}100\%$ for trans-Neptunian binaries. In large asteroid binary  systems, i.e., those in which one component has a radius exceeding \qty{20}{\kilo\meter}, the secondary-to-primary mass ratio can range from \num{e-6} to \num{e-2}~\cite{2015aste.book..355M}. 

The exchange rate for typical asteroid binary parameters is shown in case \textbf{B} of \cref{fig:exchange-rate-comparison}. Here we take $m_1 = \qty{e16}{\kilo\gram}$, $m_2 = \qty{e20}{\kilo\gram}$, and $a = \qty{e-7}{\au}$ (\qty{15}{\kilo\meter}). The exchange rate peaks for $m_\pbh \simeq m_1$, attaining a value of $\mathcal{O}(\num{e-20})\,\qty{}{\per\giga\year}$. Clearly, the exchange rate is much too small for such an event to have happened within the lifetime of the Solar System with any appreciable probability. That said, the rate is most enhanced for wide binaries with large mass ratios between the primary and secondary partner.

Setting aside the minuscule rate, the observability of an asteroid-PBH binary poses an interesting challenge. Advances in observational techniques, including radar, direct imaging, advanced lightcurve analysis, spacecraft imaging, and  adaptive optics on large telescopes have enabled the identification of large numbers of binary systems in recent years~\cite{2020tnss.book..201N, 2004Icar..168..409P, 2002aste.book..289M}. Lightcurve analysis, relevant for nonsynchronous eclipsing binaries, is likely to be ineffective, given the small radius of the PBH\@. Direct imaging is already difficult for binaries given the extremely small angular resolution required to distinguish partners separated by fractions of an arcsecond. For a PBH partner, this method becomes even more difficult. In principle, astrometric perturbations of the primary component due to the presence of a dark secondary may provide an avenue for detection in the future.

In the particular case of PBH exchange in the Solar System asteroid belt, Hawking radiation provides a reliable means of detection for light PBHs with $m_\pbh \simeq 10^{14.5}\,\qty{}{\gram}$, whose lifetimes are comparable to the age of the Universe. For a Schwarzschild BH, neglecting greybody factors, the luminosity of Hawking radiation is given by $L_\pbh = \sigma / (16\pi^{3/2}Gm_\pbh)^2$, where $\sigma$ is the Stefan-Boltzmann constant. (Here we use natural units with $c=\hbar=k_{\mathrm{B}}=1$.) Since the radiation is emitted with a blackbody spectrum, the typical energy of radiated photons is of order the temperature, $T_\pbh = (8\pi Gm_\pbh)^{-1}$, in the geometrical optics approximation. Accordingly, the flux of photons per unit area at a distance $d$ is estimated by $\Phi_\gamma \simeq L/(4\pi d^2T_\pbh)$, or
\begin{equation}
    \Phi_\gamma \simeq
    \qty{20000}{\meter^{-2}.\year^{-1}}
    \left(\frac{m_\pbh}{10^{14.5}\,\qty{}{\gram}}\right)^{-1}
    \left(\frac{d}{\qty{2}{\au}}\right)^{-2}
    .
\end{equation}
These photons would have a typical energy of $E_\gamma \sim T_\pbh \approx (\qty{30}{\mega\electronvolt})\times[m_\pbh/(10^{14.5}\,\qty{}{\gram})]^{-1}$. This estimate suggests that space-based MeV telescopes should be sensitive to the presence of such an object in the asteroid belt.

In particular, the upcoming AMEGO-X telescope~\cite{Caputo:2022xpx} has an effective area of \qty{400}{\centi\meter^2}, corresponding to a signal rate of about \qty{700}{\per\year}. Given the angular resolution of \qty{2.5}{\degree} and the energy resolution of $\mathcal O(10\%)$, the expected background rate at the spectral peak in each angular patch is $\mathcal O(\qty{e5}{\year^{-1}})$. Furthermore, this signal would exhibit a strong modulation over the period of the distance between Earth and the binary, enabling strong background rejection. As such, we expect that AMEGO-X could conclusively detect any PBH in the asteroid belt with mass of order $10^{14.5}\,\qty{}{\gram}$ within a few years.

A more refined estimate can be performed by including greybody factors using the \texttt{BlackHawk} code~\cite{Arbey:2019mbc, Arbey:2021mbl}. For a Schwarzschild BH with a mass of $10^{14.5}\,\qty{}{\gram}$, the photon spectrum peaks at \qty{190}{\mega\electronvolt}. Considering a wide energy bin of $[140,\,240]\,\qty{}{\mega\electronvolt}$, this yields a signal rate of \qty{130}{\per\year} against a background rate of \qty{9000}{\per\year}. This signal can be detected with 95\% confidence after two years of observation, even neglecting modulation. For PBHs with significant angular momentum~\cite{Arbey:2020yzj,Calza:2023iqa,Taylor:2024fvf}, the emission rate is enhanced at high energies, where the expected background rate is reduced. For a Kerr PBH with a dimensionless spin parameter $a_\ast=0.9999$ and a mass $m_\pbh=\qty{e15}{\gram}$, the photon spectrum peaks at \qty{115}{\mega\electronvolt}. In the bin $[80,\,130]\,\qty{}{\mega\electronvolt}$, the background rate is \qty{4e4}{\per\year}, while the signal rate increases to \qty{2000}{\per\year}. This high signal rate allows for the detection of PBH at the 95\% confidence level with about 15 days of observation. In principle, $\gamma$-ray signals might also be observed from PBHs transiting through the Solar System or captured via three-body processes without ejection. However, in these cases, the signal is transient. A transiting PBH is only detectable when in proximity to Earth, and a PBH captured into a triple system by three-body dynamics has a metastable orbit and will eventually be ejected. (See e.g. \refcite{Lehmann:2020yxb}.) A PBH captured into a stable orbit by ejection of a third body can remain observable for the lifetime of the Solar System, making exchanges observable now even if they occur only once in the history of the system. For a more comprehensive discussion of the observability of Hawking radiation from within the Solar System, see \refcite{Arbey:2020urq}.

\subsection{Exoplanetary systems}
\label{sec:exoplanets}
Exoplanetary systems are abundant, closely observed, and extremely well modeled. As such, it is interesting to consider the prospects for detecting exchanged PBHs in this context. If such a process were to occur in an exoplanetary system, the system would host a PBH ``planet,'' possibly in addition to other ordinary planets. The case for detecting PBHs in exoplanetary systems is ultimately similar to that in our own Solar System. Several authors have examined the possibility of PBH capture by one of our own planets~\cite{Lehmann:2022vdt}, or by other means~\cite{Scholtz:2019csj,Eroshenko:2023btr}. The relative utility of exoplanetary systems over the Solar System is twofold: first, they are numerous, widening our opportunity to find such an object; and second, the parameters of planets and their orbits span a range of values different from those in our own Solar System, possibly offering better sensitivity to certain PBH masses.

Exoplanetary systems are studied by a combination of data from Doppler spectrographs (the radial velocity method), astrometric measurements, transits, and direct imaging. Different methods have complementary strengths, and combinations of methods can yield more powerful results than any single measurement.  Astrometric measurements can reveal all the orbital properties of the exoplanet as well as its mass, but even with the levels of precision achieved by \Gaia{}, these measurements achieve only modest signal-to-noise ratios (${\sim}10$) and can only probe giant planets~\cite{2010exop.book.....S, 2022AJ....164..196W}. On the other hand, radial velocity data can yield much larger signal-to-noise ratios for giant planets, but only gives a lower limit to the exoplanet mass. Joint fitting of Doppler data and \Gaia{} astrometric data has been shown to tightly constrain the exoplanet mass~\cite{2022AJ....164..196W, 2023AJ....165..266M} and diagnose false positives such as brown dwarf companions~\cite{2023RAA....23e5022X, 2023MNRAS.526.5155S}.

Currently, \Gaia{} provides astrometric information for 73 confirmed exoplanets and candidates, of which nine have already been confirmed with Doppler data. Simulations indicate that analysis of astrometric wobble of host stars will enable \Gaia{} to discover between \num{e4} and \num{e5} exoplanets in the future~\cite{2014ApJ...797...14P}. A different combination of methods to detect exoplanets is Doppler data with transits. The NASA Exoplanet Archive~\cite{exoplanet_archive} hosts a total of 5690 exoplanets, of which 4262 have been discovered by transits, 1092 by radial velocity, and 838 by a combination of these two techniques.

The astrometric signatures of a PBH ``exoplanet'' would be identical to those of an actual planet, but a PBH of planetary mass would have an extremely small geometric cross section, and thus would not affect the lightcurve at all during a stellar transit~\cite{Bai:2023mfi}. As such, the observable effect of the presence of PBHs in exoplanetary systems would be the presence of astrometric or Doppler signatures without corresponding transit signatures. If the orbital period and phase are accurately determined through Doppler measurements, and the inclination angle is obtained from astrometric measurements, it is possible to predict the time of transit for a specific exoplanetary system. A missed transit would then indicate the presence of a PBH exoplanet. However, due to the uncertainty in measuring the orbital inclination of known exoplanetary systems, making precise transit predictions is challenging at this time. If a large fraction of exoplanets were actually PBHs, the rate of transits would be suppressed relative to the expectation from the rate of astrometric detections, providing a statistical hint for the presence of PBH exoplanets.

Thus, such a detection would only be possible if PBH planets were abundant. However, the rate of exchanges is too small for this to be realistically observable in our case. Even under the most generous assumptions, the exchange rate per binary is of order \qty{e-6}{\per\giga\year}, as shown in \cref{fig:exchange-rate-planet}. This would translate to at most $\mathcal O(1)$ PBH planet in the entire sample of exoplanets expected to be discovered in the foreseeable future. Interestingly, our computation does suggest that if planet-mass PBHs account for a large fraction of the DM, then there likely are systems with PBH planets in the Milky Way, even under pessimistic assumptions. However, since we will only be able to characterize a small fraction of these systems, a robust detection of PBHs by this method is likely impossible.

\subsection{Stellar binaries}
Stellar binaries are perhaps the most well studied population of binary systems. They are abundant and readily observable. We show exchange rates for typical stellar binary systems in \cref{fig:stellar-binary-exchange-rate}. The overall rate is only slightly higher than that in the exoplanetary case, as can be seen clearly in \cref{fig:exchange-rate-comparison}. However, the large number of detectable stellar binaries offers a distinct set of observational opportunities.

An analysis of \Gaia{} eDR3~\cite{2021A&A...649A...1G} reveals 1.2 million high-confidence binaries within \qty{1}{\kilo\parsec} of the sun with $\parallax/\sigma_\parallax > 5$ for both components, where $\parallax$ and $\sigma_\parallax$ denote the parallax and its uncertainty. Most of these are main sequence--main sequence (MS-MS) binaries with separations between a few \qty{}{\au} and $\mathcal O(\qty{e6}{\au})$~\cite{2021MNRAS.506.2269E}. A larger dataset with 1.8 million candidates becomes dominated by chance alignments at large separations.  The techniques used to identify binary systems include (i) analysis of light curves ($a \lesssim \qty{0.1}{\au}$); (ii) single- and double-lined radial velocities from the radial velocity spectrometer ($a \lesssim\textnormal{few}\,\qty{}{\au}$); (iii) astrometric perturbations such as excess astrometric noise, proper motion anomaly, or astrometric orbits ($a \lesssim \qty{10}{\au}$); and (iv) direct spatial resolution ($a \gtrsim \qty{100}{\au}$).

The identification of a BH partner in a \Gaia{} binary system  is a non-trivial question, and we mainly summarize the results given by \refcite{El-Badry:2024vjt}. The most optimistic methods rely on astrometric perturbations. Indeed, the first BHs discovered by \Gaia{} have been validated in this manner~\cite{El-Badry:2022zih, Gaia:2024ggk, El-Badry:2024isa}. Astrometric wobble has a non-trivial dependence on the mass and luminosity ratios of the binary components. Since the BH component is dark, the photocenter traces the luminous component. The  predicted angular photocenter semimajor axis for the binary is given by $\psma = q \parallax (a/\qty{}{\au})$ for $q \ll 1$ and $\psma = \parallax (a/\qty{}{\au})$ for $q \gg 1$, where $q$ is the mass ratio of the BH to the star. For a luminous companion, these relations are scaled by the light ratio: $\psma = \delta_{q\ell} \parallax (a/\qty{}{\au})$, where $\delta_{q\ell} = |q-\ell|/[(q+\ell)(\ell+1)]$ with $\ell$ being the secondary-to-primary light ratio. Detection of a BH component requires $\psma$ to be greater than the typical observation precision $\sigma_\xi$ in the along-scan direction for a given apparent magnitude.

In the aftermath of the discovery of a dark companion, the question of whether it is an astrophysical BH or a PBH becomes relevant. Binaries comprising a Solar-mass star and a sub-Solar-mass dark companion (which could plausibly be an astrophysical compact object, planet, or PBH) are difficult to observe, as discussed in \cref{sec:exoplanets}. However, \Gaia{} has already detected several binaries with a more massive dark companion~\cite{El-Badry:2022zih,El-Badry:2024isa}. In systems such as these, the dark companion can only be identified as a PBH if the properties of the binary are sufficiently anomalous to call standard astrophysical origins into question. Properties anomalous to such an origin could pertain, for example, to the mass ratio, separation, and composition being incompatible with standard isolated binary evolution models. We take up this analysis for the \Gaia{} binaries in greater detail in the following section.

\section{Exchange scenario for \Gaia{} binaries}
\label{sec:gaia}

\Gaia{} has in fact detected several binary systems with a BH component, denoted by \Gaia{} BH1, BH2, and BH3~\cite{El-Badry:2022zih, Tanikawa:2022xel,El-Badry:2024isa}. Interestingly, the properties of these binaries pose a challenge to the standard astrophysical formation channels. In this section, we pose the question of whether these binaries might have been formed by exchanges with PBHs. There are two stages to this discussion. Firstly, for a given BH-stellar binary, how likely is it that its origin is from an isolated binary system, versus a dynamical mechanism such as exchange? Secondly, if dynamical exchange is favored, how likely is it that the exchange was with an astrophysical BH versus a PBH\@? We discuss these issues in the context of the concrete  examples of \Gaia{} BH1--BH3, first summarizing the relevant features of these binaries.

\Gaia{} BH1~\cite{El-Badry:2022zih, Tanikawa:2022xel} is a binary system with a G-type main sequence star with mass $M_1 = 0.93 \pm \qty{0.05}{M_\odot}$ orbiting a BH of mass $M_2 = 9.62 \pm \qty{0.18}{M_\odot}$, first identified astrometrically and then validated spectroscopically.  The luminous component is a bright Solar-type star with $G = 13.77$ and metallicity $[\ce{Fe}/\ce{H}] = -0.20 \pm 0.05$. The semimajor axis of the binary is $a = 1.40 \pm \qty{0.01}{\au}$ with eccentricity $e = 0.451 \pm 0.005$. (We quote the central values obtained by a combination of astrometry and radial velocity.) The age of the luminous partner is at least \qty{4}{\giga\year}. The luminous G star exhibits an abundance pattern typical of the thin disk: for all measured elements \ce{X}, the abundance ratios $[\ce{X}/\ce{Fe}]$ is within $2\sigma$ of the central value for stars in the Solar neighborhood~\cite{Bensby:2013boa}. This fact, combined with the fact that the luminous star is expected to have a thin convective envelope, implies that it suffered little contamination from its partner during their possible coevolution. There is also no evidence for the contamination of the photosphere by $\alpha$-elements from the partner's death, or enrichment of $s$- and $r$-process elements. Thus, the luminous component of the binary evinces almost no evidence of a prior stage of interaction with its companion, even during the latter's putative death~\cite{El-Badry:2022zih}.

In combination, the composition of the luminous component, the mass $M_2$ of the BH, and the mutual separation of the components make it difficult to explain \Gaia{} BH1 via the evolution of an isolated stellar binary. The reasoning is as follows. The BH mass, $M_2 \approx \qty{10}{M_\odot}$, implies that the stellar progenitor had a mass 30--\qty{50}{M_\odot}, assuming Solar metallicity~\cite{Sukhbold:2015wba}, since progenitors in that range can produce $\geq \qty{9}{M_\odot}$ \ce{He} cores. A progenitor of that size  would have reached a radius of ${\sim}\qty{10}{\au}$ during its supergiant phase. Given the much smaller semimajor axis, $a \approx \qty{1.4}{\au}$, this suggests that the progenitor of the BH would have interacted  with the luminous G star in a common envelope phase. However, with the extreme mass ratio, the smaller star would not be expected to survive such a phase~\cite{Justham:2005zw}. Even allowing for the survival of the G star, the final separation following a common envelope phase would have been much smaller, ${\sim}\qty{0.02}{\au}$. This conclusion persists for  various combinations of Zero Age Main Sequence (ZAMS) properties, different models of the common envelope phase, and assumptions about the natal kick during the formation of the BH~\cite{El-Badry:2022zih}.

At present, therefore, the origin of \Gaia{} BH1 remains uncertain, with several options on the table. One possibility is that the binary was originally much more widely separated, and was significantly hardened by dynamical encounters after the formation of the BH component. Alternatively, the binary may have formed in a hierarchical triple, or other three-body system. A third possibility is a dynamical exchange process of the type described in this paper: in this scenario, one of the components was not originally part of the binary, and underwent an exchange with the original partner. If dynamical exchange is favored, the next question is whether the substituted partner had an astrophysical or primordial origin. This requires an examination of the trajectory of \Gaia{} BH1. The orbit of \Gaia{} BH1 is typical of a thin-disk star, never coming close to the Galactic Center, and limited to within $\pm\qty{250}{\parsec}$ of the disk mid-plane. Given that exchanges with astrophysical BHs are much more probable in globular clusters, the  possibility that BH1 system formed  through a dynamical process with an astrophysical BH is difficult to realize~\cite{Rastello:2023fkx}. That said, simulation results from \refcite{Rastello:2023fkx} conclude that BH1-like systems could have formed from repeated dynamical exchanges and collisions involving the progenitor star in young star clusters. \Gaia{} BH1 is similar in orbital period and mass ratio to NGC 3201 \#12560, which is located in the globular cluster NGC 3201, and for which a history of dynamical interactions (including exchange) with astrophysical BHs is much more plausible.

\Gaia{} BH1 therefore provides a typical example where processes of the kind described in our work may be fruitfully investigated.  \Gaia{} BH2 furnishes a similar example. (See \refcite{2024arXiv240313579K}, however, for a different viewpoint.) In both systems, the binary separation is small enough and the BH is massive enough that the binary components should have experienced a common envelope phase. However, such a phase would have resulted much smaller separations than actually observed. Moreover, the luminous partner shows no evidence of interacting with its companion. Isolated binary evolution is disfavored, and the binary orbit poses challenges to standard dynamical processes in astrophysical BH-rich environments as well.

\Gaia{} BH3 is an astrometric binary whose components are a \qty{0.8}{M_\odot} star and a \qty{33}{M_\odot} dark companion, with semimajor axis $a\approx\qty{16.5}{\au}$ and eccentricity $e \approx 0.73$. It presents a similar conundrum with respect to its origin as do BH1 and BH2~\cite{El-Badry:2024isa}. Its radius during a putative supergiant phase would have reached $\qty{1800}{R_\odot} \approx \qty{8.4}{\au}$, larger than the \qty{4.5}{\au} separation at periastron. Indeed, \Gaia{} BH3 presents a limiting separation beyond which evolution from a stellar binary without resorting to dynamical exchange processes becomes clearly feasible, since the BH-star system can avoid a common envelope. The \Gaia{} Collaboration has forwarded dynamical exchange as a leading alternative hypothesis for the origin of BH3~\cite{Gaia:2024ggk}. One point of difference for \Gaia{} BH3 compared to BH1 is that for the former, the metallicity is much lower than that of typical stars in the Solar neighborhood. \Gaia{} BH3 is part of the ED-2 stellar stream, and dynamical formation in the cluster that formed ED-2 has been explored by several groups~\cite{2024MNRAS.527.4031T, 2024A&A...688L...2M}.

We now turn to a discussion of the possibility that an observed star-BH binary system was formed through dynamical exchange of a star-star binary with a PBH at some stage in its history. A careful analysis would involve the following steps: for a given set of binaries, the first set of selection criteria should, of course, be those that disfavor evolution from a single isolated binary. Separations, mass ratios and compositions of the type shown by BH1--BH3 would furnish typical parameter ranges. The second set of selection criteria should maximize the probability that the binary underwent an exchange with a PBH at some stage of its history. A detailed analysis along these lines is left for the future. In the present work, we instead present a set of plausibility arguments that PBH exchange can form star-BH binaries at a non-negligible rate.

Our plausibility argument hinges on comparing two quantities: (i) the net formation rate $R_\star$ of binaries that form via a given channel (e.g., dynamical or isolated evolution) assuming the population of BHs is purely astrophysical, and (ii) the formation rate $R_\pbh$ due to exchanges with PBHs. In the context of dynamical formation, $R_\star$ is typically computed in stellar clusters as the number of binaries that form over the cluster's lifetime in ratio to its initial mass $M_\cl$. Given a formation rate per binary of $\Gamma_\star$, and assuming that the lifetime of the cluster is longer than that of a typical binary, $\tau_\bin$, we can estimate
\begin{equation}
    R_\star^{(\mathrm{ch})} \simeq
    \frac{\Gamma_\star^{(\mathrm{ch})} \times \tau_\bin \times N_\bin}{M_\cl}
    ,
\end{equation}
where $N_\bin$ is the number of binaries in the cluster that can evolve to have a BH component, and (ch) denotes the formation channel in question. Since $M_\cl / N_\bin$ is on the order of the average stellar mass, $\langle M_\star\rangle$, this rate can be estimated by $R_\star^{(\mathrm{ch})} \simeq \Gamma_\star^{(\mathrm{ch})}\tau_\bin / \langle M_\star\rangle$. Taking $\langle M_\star\rangle \sim \qty{1}{M_\odot}$, the equivalent rate for PBH exchange can then be roughly estimated by $R_\pbh \simeq \Gamma_\pbh\tau_\bin\times\qty{}{M_\odot^{-1}}$, where $\Gamma_\pbh$ is the PBH exchange rate per binary.

If these two quantities are of the same order of magnitude, then it becomes plausible that exchanges with PBHs could compare favorably with the probability of a given formation mechanism, specifically, for example, dynamical mechanisms. Of course, this does not settle the question of whether it was a PBH versus an astrophysical BH that was part of the dynamics in a given binary sample---to settle that question, a detailed study of binary orbits and stellar composition would be required to assess whether a binary passed through PBH-rich environments that were also poor in astrophysical BHs. Such an evaluation is left for future work.

Regardless of the formation mechanism, the overall rate of binary formation, $R$, can estimated observationally. Given the existence of \Gaia{} BH1, \refcite{El-Badry:2022zih} estimates that a fraction $f \sim \num{4e-7}$ of all low-mass stars should have a BH companion~\cite{El-Badry:2022zih}. This figure is obtained based on \Gaia{} DR3 data, after fitting  orbital solutions and  implementing stringent cuts: $\parallax/\sigma_\parallax > 20000 /P_{\rm orb}$ and $\psma/\sigma_{\psma} > 158/\sqrt{P_{\rm orb}}$, where $P_{\rm orb}$ is expressed in days. If \Gaia{} BH1 is taken as a prototypical binary that favors dynamical formation, then this value of $f$ would serve as a proxy for the fraction of all binaries formed in that manner. Then, since stars in the cluster have masses of $\mathcal O(\qty{1}{M_\odot})$, we have $f\sim R\times\qty{}{M_\odot}$ at the order-of-magnitude level, which in turn suggests $R\sim\qty{e-7}{M_\odot^{-1}}$. While this estimation of $f$ from~\cite{El-Badry:2022zih} is based on an analysis of observational data, a similar value, $R_\star^{\mathrm{dyn}} = \qty{2.7e-7}{M_\odot^{-1}}$, is found by \refcite{2024arXiv240313579K} from simulations of cluster models utilizing the N-body integration code \texttt{NBODY7}. Moreover, \refcite{Rastello:2023fkx} uses simulations of BH1-like binaries in low-mass and high-mass clusters at Solar metallicity, obtaining benchmark values of $R_\star^{\mathrm{dyn}} = \qty{2.09e-7}{M_\odot^{-1}}$ and \qty{2.08e-7}{M_\odot^{-1}}, respectively. 

Although isolated binary evolution is disfavored, values of $f$ for \Gaia{} BH-like binaries have also been obtained for such a formation channel. The  authors of \refcite{2024arXiv240313579K} find $R_\star^{\mathrm{iso}} \sim \qty{e-10}{M_\odot^{-1}}$ for simulations of \Gaia{} BH1 and BH2-like binaries originating via isolated binary evolution using the \texttt{StarTrack} population synthesis code, although $R_\star^{\mathrm{iso}} \sim \qty{e-7}{M_\odot^{-1}}$ can be realized if the cuts were made less stringent. Similarly, isolated binary evolution of \Gaia{} BH3-like binaries has been studied and the associated rate has been found to be $R_\star^{\mathrm{iso}} = \qty{4e-8}{M_\odot^{-1}}$ in old ($> \qty{10}{\giga\year}$) and metal-poor ($Z < 0.01$) populations~\cite{Iorio:2024pat}. We therefore arrive at the following estimates of $R$ for \Gaia{}-like BHs:
\begin{equation}
    \begin{cases}
        R_\star^{\mathrm{dyn}} \sim
            \qty{e-7}{M_\odot^{-1}}
            & \textnormal{(dynamical formation),} \\
        R_\star^{\mathrm{iso}} \sim
        \num{e-10}\textnormal{--}\qty{e-8}{M_\odot^{-1}}
            & \textnormal{(isolated binary evolution).}
    \end{cases}
\end{equation}

We now turn to an estimate of the binary formation rate by PBH exchange, i.e., $R_\pbh$. In the left panel of \cref{fig:stellar-binary-exchange-rate}, we display the exchange rate as a function of the PBH mass. \Gaia{} BH1 benchmarks are shown with the black solid line: binary component masses are \qty{1}{M_\odot} and \qty{10}{M_\odot} with separation \qty{1.4}{\au}, and the velocity dispersion is $\sigma_\dm = \qty{220}{\kilo\meter\per\second}$. Other colors show different possible choices of separations, masses, and velocity dispersions. The exchange rate increases with increasing PBH mass, reaching a maximum for $m_\pbh = \qty{1}{M_\odot}$. The rate then drops before increasing further with larger PBH mass. Neglecting bounds on the PBH abundance, the maximum exchange probability for BH1-like parameters is $\Gamma_\pbh \approx \qty{4e-9}{\per\giga\year}$. For $\tau_\bin = \qty{4.0}{\giga\year}$, we therefore obtain $R_\pbh \sim \mathcal{O}(\mathrm{few}) \times \qty{e-8}{M_\odot^{-1}}$. When accounting for observational bounds on the PBH abundance, in their most restrictive forms, this drops to $\mathcal{O}(\mathrm{few}) \times \qty{e-10}{M_\odot^{-1}}$. However, the constraints we quote in this mass range have been repeatedly disputed (see e.g. \refscite{Clesse:2017bsw,Calcino:2018mwh,Carr:2019kxo,Jedamzik:2020ypm,Jedamzik:2020omx}), meaning that this additional penalty in the rate may not apply.

However, as with astrophysical formation channels, these estimates can be significantly enhanced in stellar clusters with low velocity dispersions. For example, while globular clusters have low mass-to-light ratios, their bound DM density may still dominate over the Milky Way halo density~\cite{Carlberg:2021mpl,Mashchenko:2004hk,Shin:2013nbc,2016MNRAS.462.1937S}. In such systems, the dispersion is of order \qty{10}{\kilo\meter/\second}, corresponding to the blue curve in \cref{fig:stellar-binary-exchange-rate}. Thus, even without accounting for any DM overdensity, the exchange rate in such systems could be as high as $\Gamma_\pbh \sim \qty{e-6}{\per\giga\year}$, corresponding to a formation rate of $R_\pbh \sim \mathcal O(\mathrm{few})\times\qty{e-6}{M_\odot^{-1}}$. This rate is much larger than required to account for binaries like BH1. In fact, it is so large that PBHs could still account for BH1 even if they comprise only a small fraction of DM, as implied by the most restrictive observational constraints in this mass range. Furthermore, if PBHs account for a subcomponent of DM, they may be surrounded by a dense particle DM spike \cite{Mack:2006gz,Ricotti:2007jk,Eroshenko:2016yve}. This has been shown to enhance the rate of stellar captures \cite{Ireland:2024lye}, and may similarly influence exchange processes.

The \Gaia{} binaries BH1--BH3 themselves are not located in globular clusters, but rather likely formed in open clusters, with much lower densities. The DM density in open clusters is dynamically negligible, meaning that it must be much lower than the stellar density, but that does not preclude a bound DM component with a low velocity dispersion~\cite{Mathieu_1985}. The low density would reduce our estimated $R_\pbh$, possibly bringing it in line with observational estimates for the formation rate in open clusters. Ultimately, given the significant uncertainties in both observations and predictions, it remains plausible that BH1--BH3 and similar systems formed by dynamical exchange with PBHs, and that the BHs in these systems are primordial in origin.

\section{Discussion and conclusions}
\label{sec:discussion}

In the preceding sections, we have considered one key question: can few-body dynamics probe a population of PBHs beyond softening and disruption? While hardening is possible in principle, the effect is negligible, as we demonstrate further in \cref{sec:hardening}. PBH capture is also comparatively rare, since the resulting few-body systems are unstable to ejection. A survey of the possible outcomes of three-body encounters thus led us to study new observables associated with exchange processes, where a PBH replaces one component of a binary system. We now summarize the prospects for detecting or constraining such events.

One might hope that exchange would be viable at lighter masses than softening and disruption. The latter generally take place when PBHs are more massive than the components of the binary, which restricts their application to interesting parameter space at the planetary or asteroid mass scales. On the other hand, exchange can take place across a large range of masses, and in an enormous variety of binary systems. We therefore evaluated the rate of exchange with PBHs in several different classes of systems, including compact-object binaries, asteroid binaries, exoplanetary systems, and stellar binaries. Each of these different types of systems most readily undergoes exchanges with a different range of PBH masses, and each presents different potential observables following an exchange.

Interestingly, our findings suggest that in some of the best-motivated PBH DM scenarios, exchange processes should give rise to a sparse but nonvanishing population of PBHs in binaries in the Milky Way, including, e.g., PBH ``exoplanets.'' Moreover, for some classes of binaries, there are smoking-gun signatures of the presence of a PBH component. In particular, exchanges involving sub-Solar-mass PBHs in compact object binaries are in principle identifiable by their merger signatures, and exchanges of ultralight PBHs with asteroids in the Solar System would be observable via Hawking radiation. However, for most of these systems, we have shown that PBH exchange is unlikely to occur with a sufficient rate for any meaningful chance of observational detection.

There is one exception to this conclusion for the case of stellar binaries. Since so many of these objects are so well characterized, the low exchange rate is not an obstruction to the detection of bound PBHs. We argue that PBH exchanges in stellar binaries might be identified by the detection of binaries with BH components whose properties are otherwise difficult to explain by standard astrophysical mechanisms. Interestingly, this is exactly the case for the \Gaia{} binaries BH1--BH3: these objects are not readily accounted for by isolated binary evolution, and a dynamical origin, quite probably including exchange, is favored. We have demonstrated that the PBH exchange rate is sufficient at the order of magnitude level to produce binaries of this type in the PBH DM scenario, particularly if constraints on PBHs in the Solar mass range are relaxed compared to their most restrictive forms.

Ultimately, in order to make any robust claim of primordial origin for a stellar companion, astrophysical formation channels must be confidently excluded, and the exchange rate must be evaluated with greater specificity. The astrophysical history of a particular binary can be understood to some extent by detailed modeling of its environment and trajectory. Likewise, for a given binary, the PBH exchange rate can be accurately calibrated with numerical experiments. Thus, it is plausible that binaries like \Gaia{} BH1--BH3 will eventually furnish a detection of PBHs, not by virtue of the properties of the BHs themselves, but by virtue of the luminous objects dancing with these invisible partners.

\begin{acknowledgments}
We thank Hunter Campbell and Nathan Kaib for valuable discussions.
KS and TX thank the Center for Theoretical Underground Physics and Related Areas (CETUP*), The Institute for Underground Science at Sanford Underground Research Facility (SURF) for hospitality while this work was in progress.
BVL thanks the Kavli Institute for Theoretical Physics (KITP) for hospitality while parts of this work were completed.
The work of TX is supported by DOE grant No.~DE-SC0009956.
The work of BVL is supported by the MIT Pappalardo Fellowship.
We extend our sincere gratitude to the OU Supercomputing Center for Education and Research (OSCER) for their state-of-the-art resources that were instrumental in enabling us to conduct extensive simulations.
\end{acknowledgments}

\appendix

\section{Binary hardening by PBHs}
\label{sec:hardening}
A general gravitational interaction between a stellar binary system and a perturber results in the transfer of energy that causes changes in the mutual separation of the binary components. Generally, binaries fall into two broad categories: hard and soft. The nature of energy transfer depends on the type of binary under consideration, as discussed in \cref{sec:exchange-processes}. On an average, soft binaries gain energy from the perturber and their final orbits widen, with the system becoming less bound (softening further). On the other hand, when the binary system is hard with respect to the perturber, it typically loses energy and becomes more tightly bound. Substantial or repeated softening (for example, by MACHOs) can lead to complete disruption of binaries, and thus, the mere existence of binaries with certain parameters can be used to place constraints on the population of perturbers. The effects of hardening, on the other hand, are more challenging to observe, as we now explain.

The degree of hardening experienced by a binary system undergoes is characterized by by the hardening rate $H$, defined by the relation
\begin{equation}
    \frac{\du}{\du t}\left(\frac{1}{a}\right)
    = \frac{G\rho_{\mathrm{obj}}}{\sigma_{\mathrm{obj}}} \, H
    ,
\end{equation}
where $\rho_{\mathrm{obj}}$ is the density of the field of perturbing objects and $\sigma_{\mathrm{obj}}$ is their velocity dispersion. We now derive $H$ for a typical binary system in the presence of light PBHs with dispersion $\sigma_\dm$ and average density $\rho_\pbh$ over the binary's trajectory through the Galactic halo. The total energy of the binary is
\begin{equation}
    \label{eq:ebound}
    E_{12}
    = \frac{1}{2}\mu \vorb^2-\frac{Gm_1m_2}{r}
    =-\frac{Gm_1m_2}{2a}
    ,
\end{equation}
where $\mu$ is the reduced mass of the binary system, $m_1$ and $m_2$ are the component masses, $\vorb$ is the relative velocity of the objects in the binary, and $\sigma_\dm$ is the velocity dispersion of the perturbing PBHs. 

Now, let us assume that the binary is hard, i.e., with $v_3 \ll w(m_1, m_2)$ for $w$ the critical velocity for hardening as defined by \refcite{Quinlan:1996vp}. We also take the components to have equal mass $m_{\mathrm{c}} \equiv \frac12m_{12}$. The energy transfer in an encounter with a PBH of mass $m_3$ and velocity equal to $\sigma_\dm$ is given by
\begin{equation}
    \label{eq:hard}
    \langle \Delta E_{12} \rangle
    =-\xi \frac{m_3}{2m_{\mathrm{c}}}\frac{G m_{\mathrm{c}}^2}{2a}
    =-\xi \frac{m_3}{2m_{\mathrm{c}}}|E_{12}|,
\end{equation}
where $\xi$ is an $\mathcal O(1)$ factor determined from $N$-body simulations~\cite{1983AJ.....88.1269H}. The value of $\xi$ depends upon the initial eccentricity of the binary, the mass ratio of the binary components, and the impact parameter of the perturber. For the case of an equal-mass binary with a circular orbit, $\xi\approx 2$. From \cref{eq:hard}, we can determine the rate at which the binary loses energy over many encounters. Since the number density of PBHs is $n_\pbh = \rho_\pbh / m_3$, we have
\begin{equation}
    \label{eq:hardening}
    \frac{\du E_{12}}{\du t}
    = \langle \Delta E_{12} \rangle\frac{\du N}{\du t}
    = -\pi\xi(Gm_{\mathrm{c}})^2\frac{\rho_\pbh}{\sigma_\dm}
    .
\end{equation}
Here, the encounter rate is given by $\du N/\du t = n_\pbh\sigma_\dm\Sigma_3$, where $\Sigma_3$ is the cross section for three-body encounters. This cross section is estimated using gravitational focusing and is given by
\begin{equation}
    \Sigma_3 = \frac{2\pi Gm_{123}a}{\sigma_\dm^2}
    .
\end{equation}
We can rewrite \cref{eq:hardening} in terms of $a$ as
\begin{equation}
    \label{eq:dadt}
    \frac{\du}{\du t}\left( \frac{1}{a} \right)
    = \frac{G\rho_\pbh}{\sigma_\dm}H_0
    ,
\end{equation}
so $H_0 = 2 \pi \xi$ is the hardening rate induced by slow PBHs with velocity dispersion $\sigma_\dm \ll w$. This is not appropriate for a realistic PBH distribution: $w \simeq 0.6\times\vorb$ for an equal-mass binary, typically well below $\sigma_\dm$. At higher velocities, \cref{eq:dadt} needs to be modified. Following \refcite{Quinlan:1996vp}, we define the velocity-dependent hardening rate as
\begin{equation}
    H_1 \left( v_3 \right)
    = \frac{H_0}{\left[ 1 + (v_3/w)^4 \right]^{1/2}}
    .
\end{equation}
As the velocity of the PBH increases, the hardening rate decreases and goes as $H_0/v^2_3$ for $v_3 \gg w$. For a Maxwellian velocity distribution, we finally obtain 
\begin{equation}
    \label{eq:hardening-rate}
    H
    = \int_0^\infty\du v_3 \, \fdm(v_3)
    \frac{\sigma_\dm}{v_3} \,  H_1(v_3)
    ,
\end{equation}
where $\fdm(v_3)$ is defined in \cref{eq:vdf}.

\Cref{eq:hardening-rate} gives the appropriate hardening rate for a binary system in a homogeneous environment of low-mass PBHs with a velocity dispersion typical of DM\@. However, the hardening rate is dominated by perturbers with $v_3$ not much greater than $w$, for which $H_1$ is approximately independent of $a$. In this case, \cref{eq:dadt} gives $\du a/\du t \propto a^2$, and the semimajor axis thus evolves at a characteristic rate
\begin{equation}
    \left|\frac{1}{a}{\frac{\du a}{\du t}}\right| \simeq
    \frac{2\pi\xi G\rho_\pbh a}{\sigma_\dm}
    \approx \qty{0.01}{\giga\year^{-1}}
    \left(\frac{a}{\qty{e6}{\au}}\right)
    ,
\end{equation}
where we take fiducial values $\xi=2$, $\sigma_\dm = \qty{220}{\kilo\meter/\second}$, and $\rho_\pbh = \qty{0.4}{\giga\electronvolt/\centi\meter^{3}}$. This estimate already indicates that hardening due to PBHs is potentially observable only for the widest binaries.

Furthermore, while these light PBHs harden a given binary, that same binary is subject to softening due to encounters with ordinary stellar objects when it transits through the Galactic disk. The change in energy of the binary per unit time due to multiple encounters with stellar objects is~\cite{2008gady.book.....B}
\begin{equation}
    \label{eq:softening}
    \frac{\du E_{12}}{\du t}
    = 8\sqrt{\frac{\pi}{3}}
    \frac{G^2m_{\mathrm{c}}m_\star \rho_\star}{\sigma_\star}
    \log\Lambda
    ,
\end{equation}
where $m_\star$ is the mass of each stellar object, $\rho_\star$ is their average density, and $\sigma_\star$ is their velocity dispersion. Here $\log\Lambda = \log({b_{\max}}/{b_{\min}})$ is the Coulomb logarithm, with $b_{\min}$ and $b_{\max}$ denoting the minimal and maximal impact parameters, respectively. We fix $b_{\min} = G(m_{\mathrm{c}} + m_\star)/\langle v_\star^2\rangle$, where $v_\star$ is the relative velocity between stars and the center of mass of the binary, and we approximate $\langle v_\star^2\rangle = (2.1\sigma_\star)^2$~\cite{1987gady.book.....B}. To ensure \cref{eq:softening} is valid, the impact parameter should be much smaller than the binary star separation, such that the change of velocity of the distant binary component is negligible. Therefore we set the $b_{\max}=a/2$, and we include a factor of $2$ to account for interactions with either of the binary components.

The softening rate $S$ due to stellar encounters is defined similarly to the hardening rate:
\begin{equation}
    \label{eq:softening-definition}
    \frac{\du}{\du t} \left( \frac{1}{a} \right)
    = -\frac{G\rho_\star}{\sigma_\star}S
    .
\end{equation}
From \cref{eq:softening-definition}, we identify the softening rate as $S = 16\sqrt{\pi/3}\log\Lambda$, with $\Lambda = 2.2a\sigma_\star^2/[G(m_{\mathrm{c}} + m_\star)]$. For simplicity, we now assume that all stars have the same mass as the binary components, $m_\star = m_{\mathrm{c}}$. To minimize the softening, we also assume that the trajectory of the binary is such that it is outside of the Galactic disk for much of its orbit, spending a fractional time $x$ in a star-rich environment. Neglecting the weak dependence of softening on $a$, and given some initial separation $a_0$, the separation of a typical binary after a time $t$ will be determined by a combination of softening and hardening effects:
\begin{equation}
    \label{eq:evolved-separation}
    a(t) = \left[
        a_0^{-1} + 
        \frac{G\rho_\pbh}{\sigma_\dm} Ht
        - \frac{G\rho_\star}{\sigma_\star}Sxt
    \right]^{-1}
    .
\end{equation}
The impact of hardening by low-mass PBHs is only non-negligible if the hardening term in \cref{eq:evolved-separation} is at least as large as the softening term. For a binary with a semimajor axis of \qty{e5}{\au}, we find that in order for the hardening term to exceed the softening term, we must have
\begin{multline}
    H \gtrsim 900 \left(\frac{\sigma_\dm}{\qty{220}{\kilo\meter/\second}}\right)
    \left(\frac{\rho_\pbh}{\qty{0.4}{\giga\electronvolt/\centi\meter^{3}}}\right)^{-1}
    \\\times
    \left(\frac{x}{0.1}\right)
    \left(\frac{\rho_\star}{\qty{2.2}{\giga\electronvolt/\centi\meter^{3}}}\right)
    \left(\frac{\sigma_\star}{\qty{25}{\kilo\meter/\second}}\right)^{-1}
    ,
\end{multline}
where we take the stellar density from \refcite{Xiang:2018azk, 2021ApJ...916..112N} and the stellar velocity dispersion from \refcite{Rix:2013bi}. However, for the same parameter values, \cref{eq:hardening-rate} implies a much smaller value of $H=\num{1.1e-5}$. Therefore, in the Milky Way, softening always dominates over hardening due to the high velocity dispersion of DM, making hardening from low-mass PBHs extremely difficult to study.

\bibliography{references}

\end{document}